\documentclass[fleqn,10pt]{wlscirep}
\usepackage[utf8]{inputenc}
\usepackage[T1]{fontenc}
\usepackage{siunitx}
\usepackage{comment}
\usepackage{graphicx}
\usepackage[export]{adjustbox}
\usepackage{subcaption}
\usepackage{caption}
\usepackage{subcaption}

\title{A Silicon Photonic 32-Input Coherent Combiner for Turbulence Mitigation in Free Space Optics Links}

\author[1,*]{Lorenzo De Marinis}
\author[1]{Peter Seigo Kincaid}
\author[2]{Yann Lucas}
\author[2]{Lea Krafft}
\author[2]{Vincent Michau}
\author[3,+]{Matteo Cherchi}
\author[3]{Mikko Karppinen}
\author[1]{Giampiero Contestabile}
\affil[1]{Scuola Superiore Sant'Anna, Pisa, 56124, Italy}
\affil[2]{DOTA, ONERA, Université de Paris-Saclay 92320, Châtillon, France}
\affil[3]{VTT Technical Research Centre of Finland, Espoo 02044, Finland}

\affil[*]{lorenzo.demarinis@santannapisa.it}

\affil[+]{Now with Xanadu Quantum Technologies Inc., Toronto, Canada}


\begin{abstract}
A photonic integrated circuit (PIC) for the coherent combination of 32 input optical signals into a single output fiber is reported. The PIC was fabricated using a low-loss thick silicon-on-insulator (SOI) process and packaged with 32 input and 1 output fibers. The basic building block is a 2x2 Mach-Zehnder interferometer (MZI) with an external (to the MZI branches) and an internal thermal phase shifter, and a bandwidth in excess of 80 kHz. The PIC monolithically integrates 31 MZIs and 31 germanium photodetectors, and is suitable in principle for turbulence mitigation in LEO-ground and horizontal free space optics links. Improvements to the device for the coherent combination of 64 inputs and for the reduction of insertion losses are also discussed.
\end{abstract}
\begin{document}

\flushbottom
\maketitle
\thispagestyle{empty}

%\noindent Please note: Abbreviations should be introduced at the first mention in the main text – no abbreviations lists. Suggested structure of main text (not enforced) is provided below.

\section*{Introduction}

Free space optical communications (FSO) have been the subject of active development for several decades. FSO links offer the potential for data rates currently inaccessible with radio frequencies, with experimental systems able to sustain terabits/s data rates over km distances~\cite{PoliakDemonstration:22}. Generally speaking, in all situations where laying optical fibers is either too expensive or not feasible, FSO can provide a viable solution~\cite{Gailani2021FSO}, and is expected to play a crucial role as one of the technologies enabling 5G/6G~\cite{Dang_Amin_Shihada_Alouini_2020} and satellite communications~\cite{Toyoshima:21}, including quantum links~\cite{liao2017satellite}. Furthermore, the short wavelength allows beams of high directivity which therefore limits possible interference problems and reduces the power required for transmission. The choice of the fiber optics C-band and the coupling of the free space beam to a single mode fiber (SMF) allow the use of existing fiber optic communication technologies and systems, thereby limiting deployment costs. However, the propagation of the beam through turbulent atmosphere is detrimental to the SMF coupling efficiency. Adaptive optics is envisaged for turbulence mitigation in future optical ground stations dedicated to ground-satellite links (GEO or LEO) to maximize the detected signal allowing throughputs of 10 to 100 Gbps per channel~\cite{fischer_esa_2019,kolev_status_2022,roberts_performance_2023,petit_oneras_2023}, some demonstrations of horizontal optical links have also been carried out~\cite{stotts_optical_2009,li_slant-path_2017}, with a transmission capacity exceeding 1~Tbps~\cite{ciaramella20091}.

\begin{figure}[t]
\centering\includegraphics[width=0.85\textwidth]{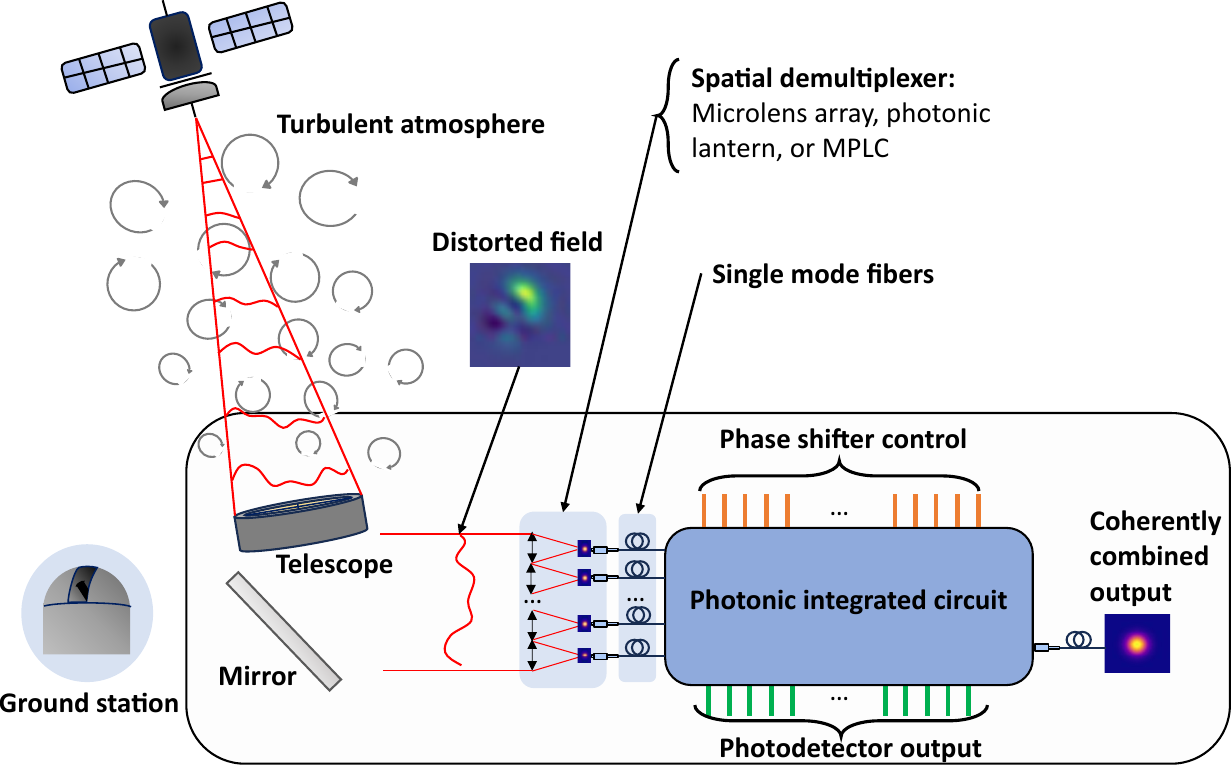}
\caption{Satellite-to-ground FSO communication with distorted fields, modal decomposition followed by PIC coherent combination after FSO beam reception by the ground station telescope.}
\label{fig:FSO}
\end{figure}

Classic adaptive optics methods for turbulence mitigation require a deformable mirror and a wavefront sensor, with relatively large footprint. Furthermore, wavefront sensors are well suited to astronomical observation and to optical links close to the zenith, but are sensitive to the strong scintillation encountered when the beams propagate in the lower atmosphere over distances greater than a few kilometers~\cite{primmerman_atmospheric-compensation_1995}. An alternative solution based on modal decomposition and subsequent guided wave coherent combination was proposed for SMF coupling~\cite{schwartz2009precompensation}; Fig~\ref{fig:FSO} illustrates the principle in the case of a satellite to ground FSO link with beam propagation through a turbulent atmosphere. Here, the distorted field is first decomposed into N guided waves via a spatial demultiplexer, which are then added coherently within a photonic circuit whose phase shifters correct the fluctuations in the complex amplitudes of the guided waves in real time. In this scheme, the phase shifters are controlled in a closed-loop, with the correction to be applied estimated from the output signals by adding a rapid, low amplitude modulation to the phase shifter control signals\cite{omeara_multidither_1977,shay_first_2006,ma_coherent_2010}. Finally, the constructively recombined output of the photonic circuit is coupled to an output SMF. Performing the coherent combination with bulk fiber components is impractical because of large size, the requirement of matched optical paths and for mechanical stability issues. For these reasons, the use of Photonic Integrated Circuits (PIC) was proposed~\cite{schwartz2009precompensation}, which offer small footprint, mechanical and thermal stability, and low mismatch between optical paths~\cite{Billault:21,Billaud2022Turbulence,melloni2023programmable}. This solution allows to overcome both the disadvantages of the large footprint and the susceptibility to strong scintillation associated with classic adaptive optics assemblies. An example of a photonic circuit architecture for the coherent addition of guided waves is presented in Fig.~\ref{fig:schematics}.

The practical specifications of coherent combiners may be deduced as a first approximation from that of the equivalent classic adaptive optics system. The performance depends firstly on the number of spatial modes, i.e., the number of actuators in the deformable mirror or the number of guided waves in the spatial multiplexer. The closed-loop correction frequency also plays an important role for both devices; when using a PIC, aside from the correction signal, an additional low amplitude tone must be applied to the phase shifters, with its relative frequency as high as possible to avoid any interference with the control signal. %\cvm{The performance depends firstly on the number of spatial modes of correction, i.e. the number of actuators in the deformable mirror or the number of guided waves in the spatial multiplexer. }
%\cvm{Performance also depends on the closed-loop correction frequency  for both devices. However, in the case of PIC, an additional  modulation has to be applied to the phase shifters, with a  frequency as high as possible with regards to the correction frequency to avoid any coupling.}
%The bandwidth of the phase shifters is determined by the maximum modulation frequency of the phase shifters which must be an order of magnitude higher than the correction frequency of the classic adaptive optics 
According to N. Védrenne~\cite{vedrenne_adaptive_2016}, a few dozen actuators would be required for AO dedicated to LEO-ground links, with the number increasing to 100-200 for GEO-ground links; a correction frequency of between 2 and 10~kHz would be required for LEO-ground links due to the movement of the satellite, this is 10 times lower for GEO-ground links. Finally, the main interest of the correction lies in the increase in power of the collected signal, hence the losses in the component must be minimized. For comparison, the typical optical losses of classic adaptive optics systems are of the order of 5~dB, due to the presence of a few dozen optical surfaces in the beam path and SMF coupling. 

%\cvm{The next sentence is not about AO requirements, but rather about PIC technology. I suggest including this sentence in the following paragraph}

In the literature, previous coherent combination schemes were demonstrated with up to 16 inputs on a single PIC fabricated with SOI~\cite{Billaud2022Turbulence, Billault:21,melloni2023programmable}, and up to 32 effective inputs by combining four 8 input PICs~\cite{Michel100Gbps2023}. Recently, the fabrication of a 45 input PIC for turbulence mitigation was reported, but without integrating the photodetectors~\cite{Billaud45mode2024}. In this paper, we report the design, fabrication, and characterization of an integrated coherent combiner suitable for 32 inputs/modes. As the scalability and footprint of the device is governed by losses and integration density, we fabricated the PIC using the low-loss silicon thick SOI platform available at VTT Finland~\cite{aalto2019open}. The 32 input configuration with phase shifter bandwidth in excess of 50 kHz, is suitable in principle for correcting the effects of atmospheric turbulence on any LEO-ground link. Indeed, these parameters are close to those of the AOptix R3.1 classic adaptive optics element used as part of the ORCA (Optical RF Communications Adjunct)~\cite{young2015development}. The fabricated chip demonstrated a bandwidth larger than 80~kHz, with phase shifters able to provide more than 6$\pi$ of phase shift correction. The integrated germanium photodetectors exhibited a responsivity of 0.8~A/W, a dark current <~2~nA, and a bandwidth >~2~GHz with zero bias applied. The remainder of the paper is organized as follows. First, we discuss the features of photonic integration platforms and outline their shortcomings for the development of coherent combiners. We then report the design, fabrication and packaging of the coherent combiner. In the device characterization section, we report the main characterization measurements of the PIC and its building blocks: phase shifters, Mach-Zehnder interferometers, photodetectors, and I/O losses. This is followed by a discussion section on our findings and prospected devices based on the same  technology, exploring their applications in realistic FSO link. Finally, we summarise the work in the conclusions.

\section*{Coherent Combiner for Free Space Optics}
\label{sec:combiner}

There are several photonic integration platforms suitable for the manufacturing of coherent combiners. Each one with its own characteristics, strengths, and limitations. However, the need of scaling the structure towards a high number of inputs poses a stringent requirement in terms of compactness: the larger the number of inputs, the larger the number of combining elements (MZIs) to be integrated and thus the associated device footprint. This renders a fabrication with the Indium Phosphide technology unpractical, due to the large bending radius and typical attainable yields~\cite{Dema2023Analysis}. The emerging lithium niobate on insulator technology (LNOI) platform provides very low losses and medium-short bending radius ($\leq$100 \si{\micro\metre}). However, its lack of maturity poses limits to practical usage: just one company in the world currently offers wafer-scale fabrication, furthermore LNOI phase shifters notoriously suffer from DC drifts, it lacks integrated PDs, and the optical I/O with fiber arrays has low yield and high losses. The Silicon Nitride platform offers a more robust process, with low coupling and propagation losses, and the possibility to hybridly integrate photodetectors. However, the scaling of structures in silicon nitride is hindered by the high bending radius of the waveguides, and the large power consumption (and related thermal crosstalk) of the thermal phase controllers.

Conversely, silicon photonics meets all the requirements as it offers compact structures (\si{\micro\metre} bends) and integrated detectors, while providing high yields with a robust process which inherited the maturity of the CMOS technology~\cite{siew2021review}. Among the various silicon platforms, the thick SOI process available at VTT was ideal for fabricating the coherent combiner demonstrator. This fabrication platform is based on 3~\si{\micro\metre} thick waveguides where the optical field modes are almost completely confined in the silicon core. This gives small propagation losses ($\sim$0.1~dB/cm in the C-band), low polarization dependency, and dense integration thanks to Euler bends that can provide an effective banding radius as small as 1.3~\si{\micro\metre}~\cite{aalto2019open}. Optical I/O is facilitated by the horizontal taperings of waveguides on chip, from 3~\si{\micro\metre} to 12~\si{\micro\metre}. The tapers work as spot-size converters to improve the coupling to SMFs. In addition, VTT offers an SOI optical waveguide interposer chip as a vertical spot-size converter to further increase the coupling efficiency between fiber arrays and the SOI PIC~\cite{aalto2019open}. Using the optical interposer, approximately 1.5-2 dB I/O coupling losses are possible, depending on the detailed structure of the I/O facet. Moreover, the available integrated germanium photodiodes  have a low dark current and a responsivity of $\approx$ 0.8 A/W.

%Optical I/O can be performed by interfacing lensed fibers with 2.5~\si{\micro\metre} spot size to the 3~\si{\micro\metre} waveguides, with a coupling loss as low as 1-1.5~dB. Alternatively, an optical interposer can be used to interface a fiber array with the chip. The integrated germanium photodiodes are expected to have a responsivity of $\approx$ 0.8 A/W. 

\begin{figure}[t]
        \centering
    \begin{subfigure}[b]{0.495\textwidth}
        \centering
        \includegraphics[width=\textwidth]{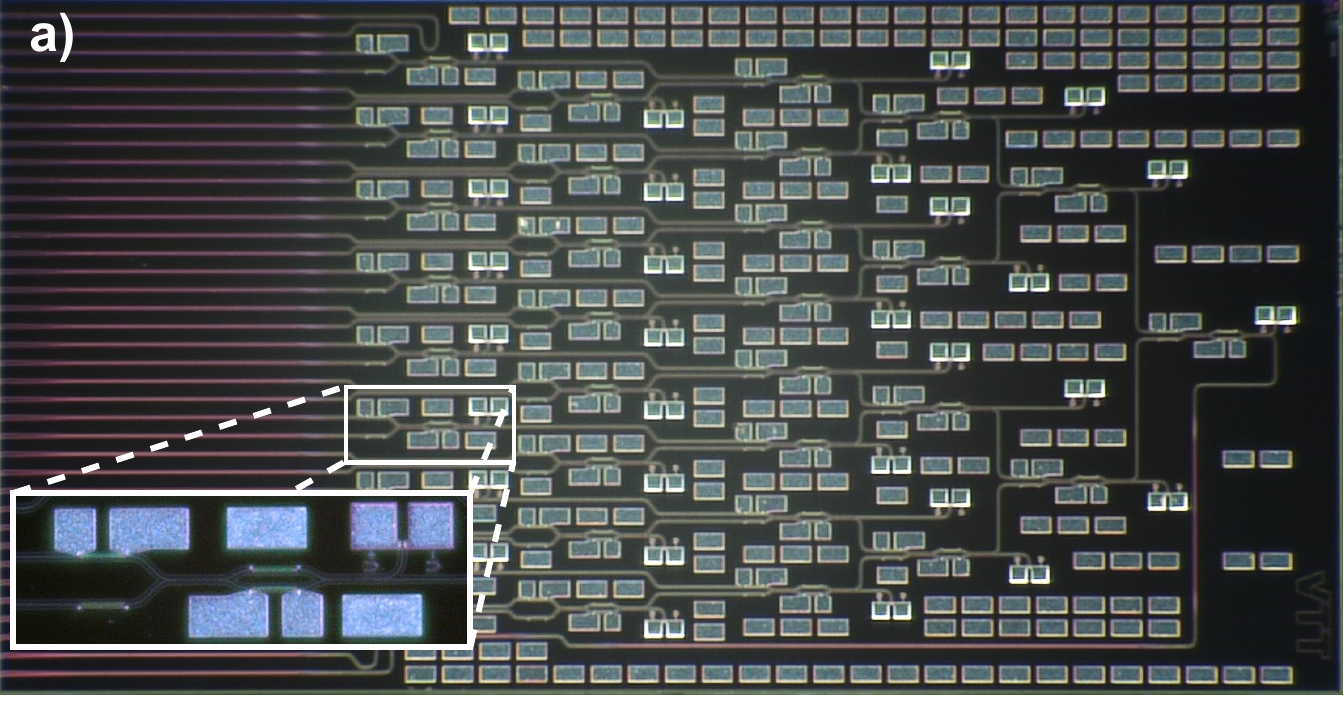}
        \phantomcaption
        \label{fig:schematic1}
    \end{subfigure}
    \hfill
    \begin{subfigure}[b]{0.495\textwidth}
        \centering
        \includegraphics[width=\textwidth]{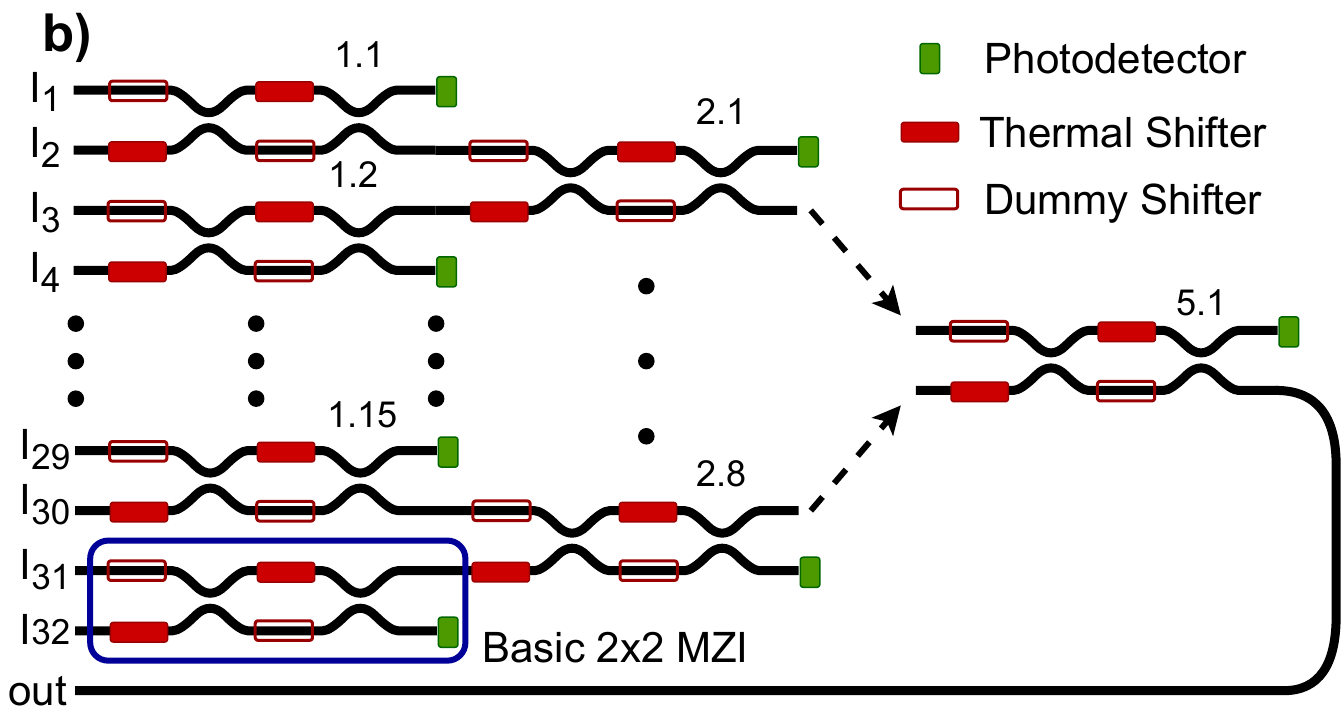}
        \phantomcaption
        \label{fig:schematic2}
     \end{subfigure}
    \caption{a) Photo of the as-fabricated Thick SOI chip with a magnification of the basic MZI; b) scheme of the coherent combiner architecture}
    \label{fig:schematics}
\end{figure}

When scaling structures that rely on thermo-optic phase shifters, assuring a high thermal isolation is of paramount importance for correct functionality. However, high integration density with close structures often leads to poor thermal isolation with an evident footprint-isolation trade-off to be made. Nonetheless, the VTT process tackles this issue: the thermal phase shifters are made of doped silicon implanted close to the optical waveguides for high efficiency, with isolation trenches that prevent thermal cross-talk between close structures. The heaters have a power consumption associated with a $\pi$ shift in phase (P$_\pi$) lower than 40~mW, and a bandwidth of several tens of kHz, making them suitable for the turbulence mitigation application. Moreover, the heaters allow for several $\pi$ of phase shift (on the order of 5-6$\pi$ before failure), a feature needed to reduce the fading induced by an abrupt $2\pi$ phase shift from the adaptive optics correction algorithm, which occurs when the upper limit in phase shift is reached.

Leveraging the features of the thick-SOI process, we were able to design an integrated coherent combiner with 32-inputs. The combination is performed with a binary tree architecture, where 2$\times$2 MZIs are used to coherently combine lightwave pairs~\cite{Miller:13}. Figure~\ref{fig:schematic1} reports a photo of the naked fabricated PIC, with the enclosed close up displaying the basic 2$\times$2 MZI element. The MZI structure consists of two inputs, one of which has a thermal phase shifter (the ``external'' shifter), a 2$\times$2 3-dB multimode interferometer (MMI) which splits evenly to two internal branches, one with another phase shifter (the ``internal'' shifter). Another MMI follows with one output coupled to a monitor photodiode (PD), and the other which feeds into the next MZI or, in the case of the last MZI, to the output port. The external phase shifter sets the correct relative phase between the two incoming lightwaves such that the power in both internal branches is equal, the internal phase shifter then directs all optical power to the correct exit. The power at the exit of each MZI may be optimised using a routine which minimises the power at the monitor PDs by opportune control of the two phase shifters. The MMI should be well balanced for a MZI to be able to combine channels with different phase and amplitude. Futhermore, optical cross talk needs to be minimal for the modulation based control algorithm not to fail. Figure~\ref{fig:schematic2} shows the equivalent schematic of the whole structure, highlighting the basic 2$\times$2 MZIs with a blue rectangle; to ensure maximally balanced paths within the MZIs, some non-connected, or ``dummy'', phase shifters were placed. The MZIs are numbered in an ascending fashion following columns and rows, e.g., MZI 1.15 is the 15$^{th}$ from the top of the PIC in the first column and MZI 3.1 is the topmost of the third column. 32 polarization maintaining single mode fibers (SMFs) are edge-coupled to the side of the chip to provide the optical I/O, the inputs feed into the first row of MZIs (1.1-1.16). The optical connections within (and between) the MZIs are strip waveguides with a height of \SI{3}{\micro\metre} and a width of \SI{1.875}{\micro\metre}, which ensure high mode confinement and small bending radius, but in which higher order modes are supported. To remove any optical power coupled to higher order modes, input lightwaves passes through a \SI{2}{\milli\metre} long section of straight rib waveguides, before coupling to the MZIs. This section can be made significantly shorter, but has been kept as long as possible considering the footprint of our chip with minimal optical losses associated with the thick SOI waveguides. The silicon chip contains a total of 31 MZIs, 62 thermal shifters, and 31 PDs, within a footprint of 10$\times$5~\SI{}{\square\milli\metre}.

%The MZI structure consists of two inputs, with a phase shifter on one of the input branches (the ``external'' shifter), a 3-dB multimode interferometer (MMI) which splits to two internal branches, one with a phase shifter (``internal'' shifter). Another MMI follows with one output coupled to a monitor photodiode, and the other which feeds into the next MZI or, in the case of the last MZI, to the output port. The external phase shifter ensures that in each of the internal branches the waves have the same magnitude and thus that a good extinction ratio can be achieved, the internal phase shifter aligns the phases of the two waves, ensuring maximum power at the correct exit. The power at the exit of each MZI may be optimised using a routine which minimises the power at the monitor PDs by opportune control of the phase shifters. On the opposite branch to the active phase shifters, dummy phase shifters without electrical connections were placed in order to balance the losses. In figure \ref{Fig:Schematic}\textbf{a} a photo of the fabricated PIC is presented, with the enclosed close up displaying the individual 2X2 MZI structure, figure \ref{Fig:Schematic}\textbf{b} shows a schematic of the same structure and labelling scheme. Polarization maintaining single mode fibers (SMFs) are edge-coupled to the side of the chip to provide the optical IO, the inputs feed into the first row of MZIs (1.1-1.16) for coherent combination.

\begin{figure}[t]
     \centering
    \begin{subfigure}[b]{0.495\textwidth}
        \centering
        \includegraphics[width=\textwidth]{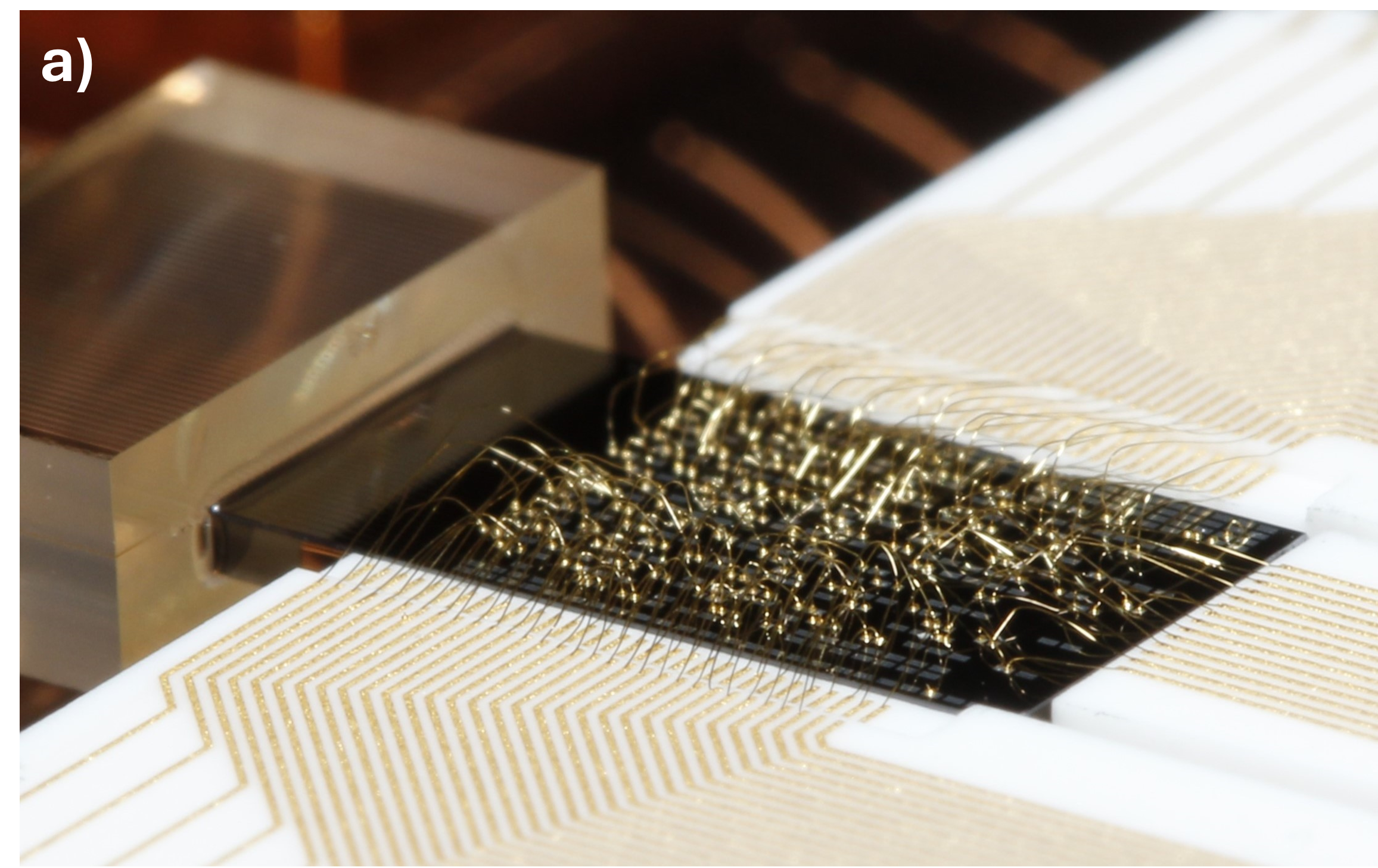}
        \phantomcaption
        \label{fig:packaged1}
    \end{subfigure}
    \hfill
    \begin{subfigure}[b]{0.495\textwidth}
        \centering
        \includegraphics[width=\textwidth]{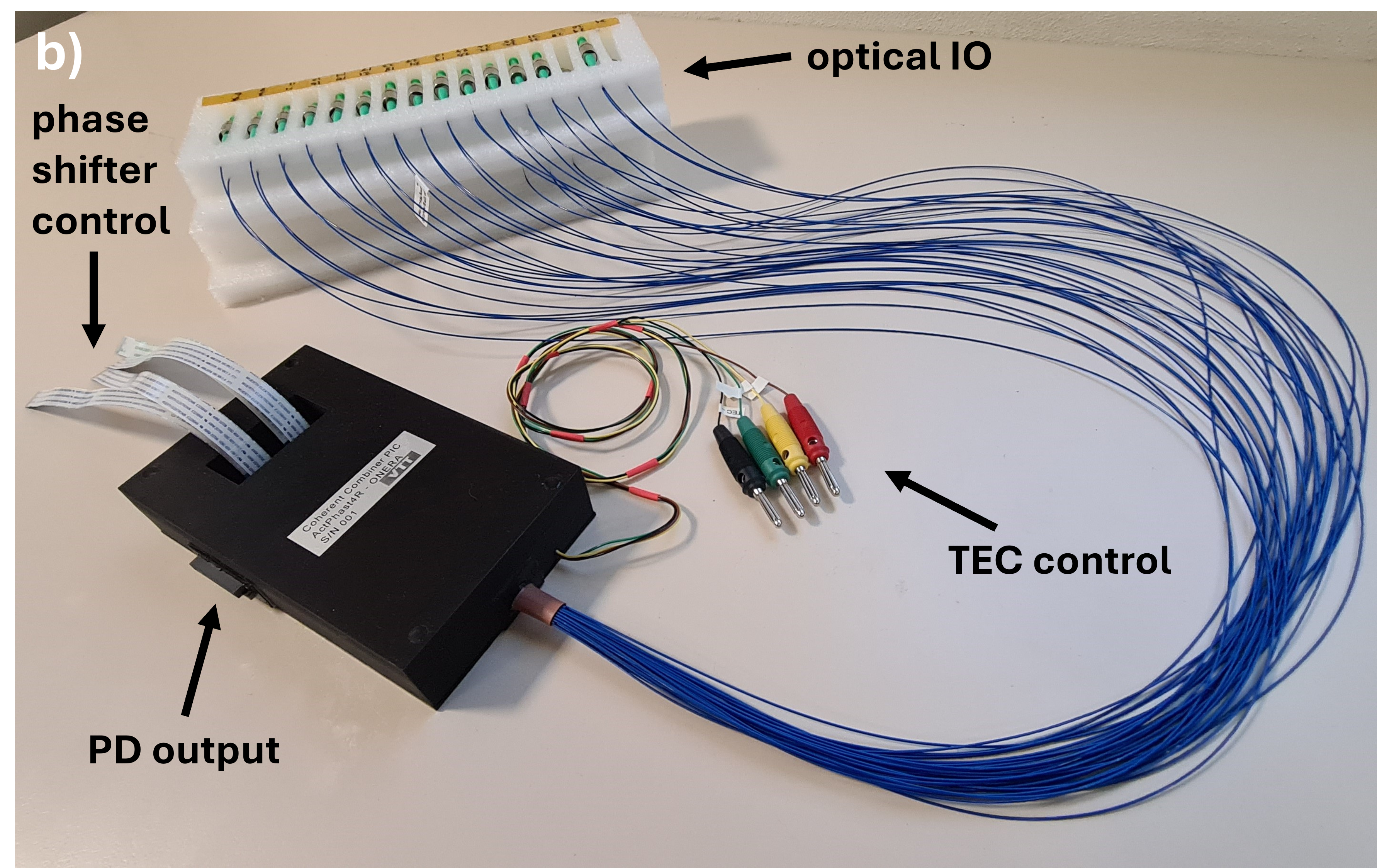}
        \phantomcaption
        \label{fig:packaged2}
     \end{subfigure}
    \caption{a) Photo of the PIC undergoing packaging after wire bonding to an electrical interposer and fixing the optical IO to a fiber array; b) Finished package with optical IO coupled to SMF fibers, thermo-electric control (coloured cables) and DC IO ports for thermal shifter control and photodiode output.}
    \label{fig:Packagedtotal}
    
\end{figure}

The chip was packaged by VTT, the electrical pads were wire bonded to a ceramic electrical interposer, and then to a PCB, and it was enclosed with a thermoelectric cooler (TEC) element in an aluminium case to allow thermal dissipation. For technical issues, it was not possible to include the optical interposer in the packaged prototype device. This because of small positioning error of the lane etching step which defines the optical I/O facet of the PIC. This error prevented the attachment of the interposer chip close enough to the waveguide facet for efficient optical coupling. Figure~\ref{fig:packaged1} shows the chip after wirebonding and attachment of the optical fiber array, figure~\ref{fig:packaged2} shows the fully-packaged chip. The coloured DC cable connections are for piloting the TEC, on the case the laterally placed DC connectors are for the monitor photodiodes, and the central DC connectors are for MZI control.

\section*{Device Characterization}
\label{sec:char}

%The Discussion should be succinct and must not contain subheadings.
%\cpk{introduction, suggest writing by Onera, to justify some of the tests in context of the combiner, need the first figure in the paper to illustrate the concept/application, ask onera}

The packaged chip was characterised with an external cavity tunable laser, individually coupled to the optical inputs. The MZIs were driven with a custom digital to analog converter (DAC), whose output ranges between 0 and 32~V. The output port was connected to a PD for analysis. The temperature of the TEC was set to 20~$^{\circ}$C unless otherwise specified. Information about the equipment used, the experimental setup and additional characterizations may be found in the supplementary information.

%The packaged chip was characterised with an external cavity laser with tunable wavelength, individually coupled to the optical inputs. A custom digital to analog converter (DAC), with output range 0-32 V, interfaced with the MZIs in order to manipulate the routing of the light. The output port was connected to a PD for analysis. The temperature of the TEC was set to 20~$^{\circ}$C unless otherwise specified.

\begin{figure}[h]
    \centering
    \begin{subfigure}[b]{0.49\textwidth}
        \centering
        \includegraphics[width=\textwidth]{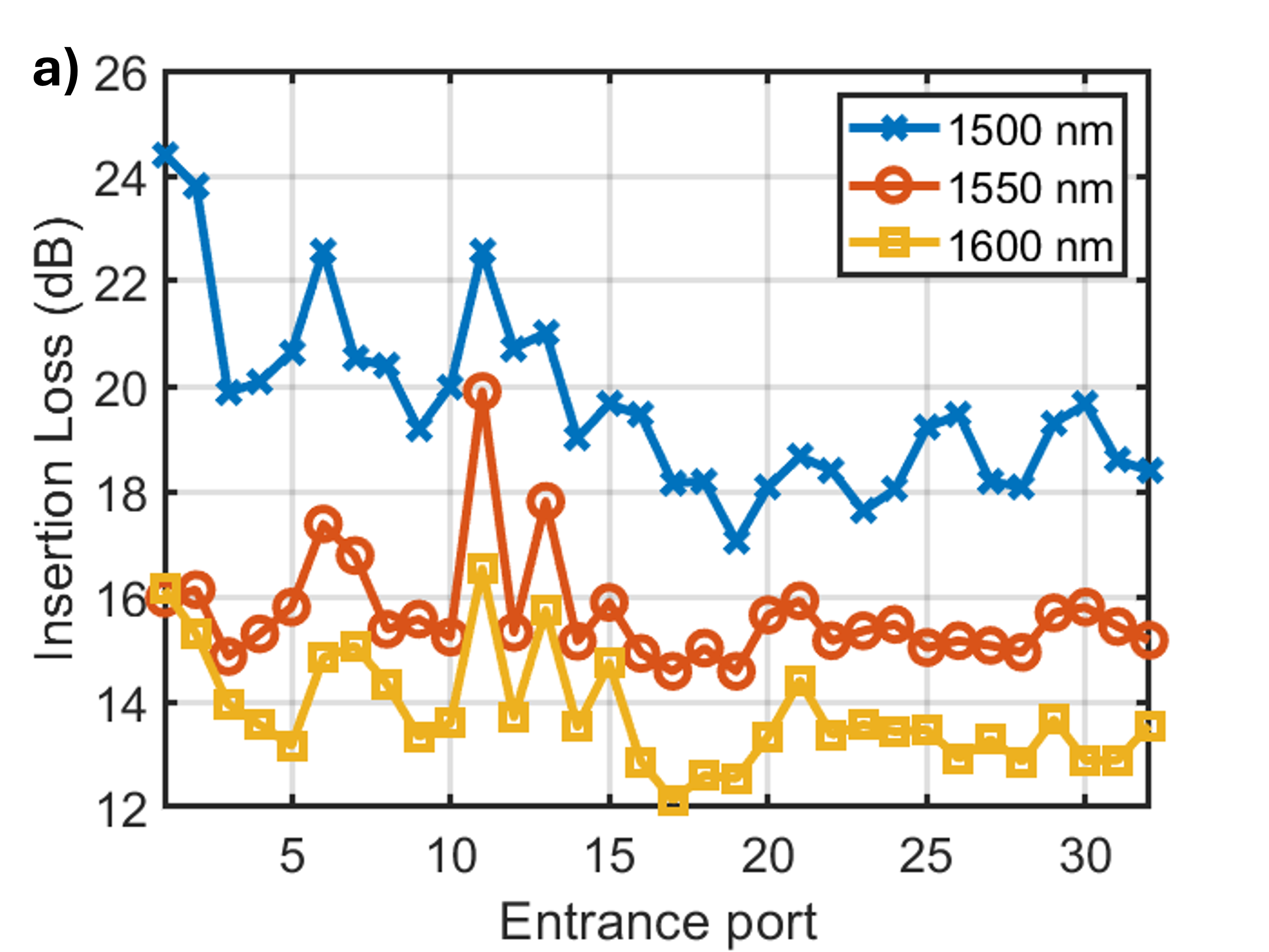}
        \phantomcaption
        \label{fig:MZM_Graphs_left}
    \end{subfigure}
    \hfill
    \begin{subfigure}[b]{0.49\textwidth}
        \centering
        \includegraphics[width=\textwidth]{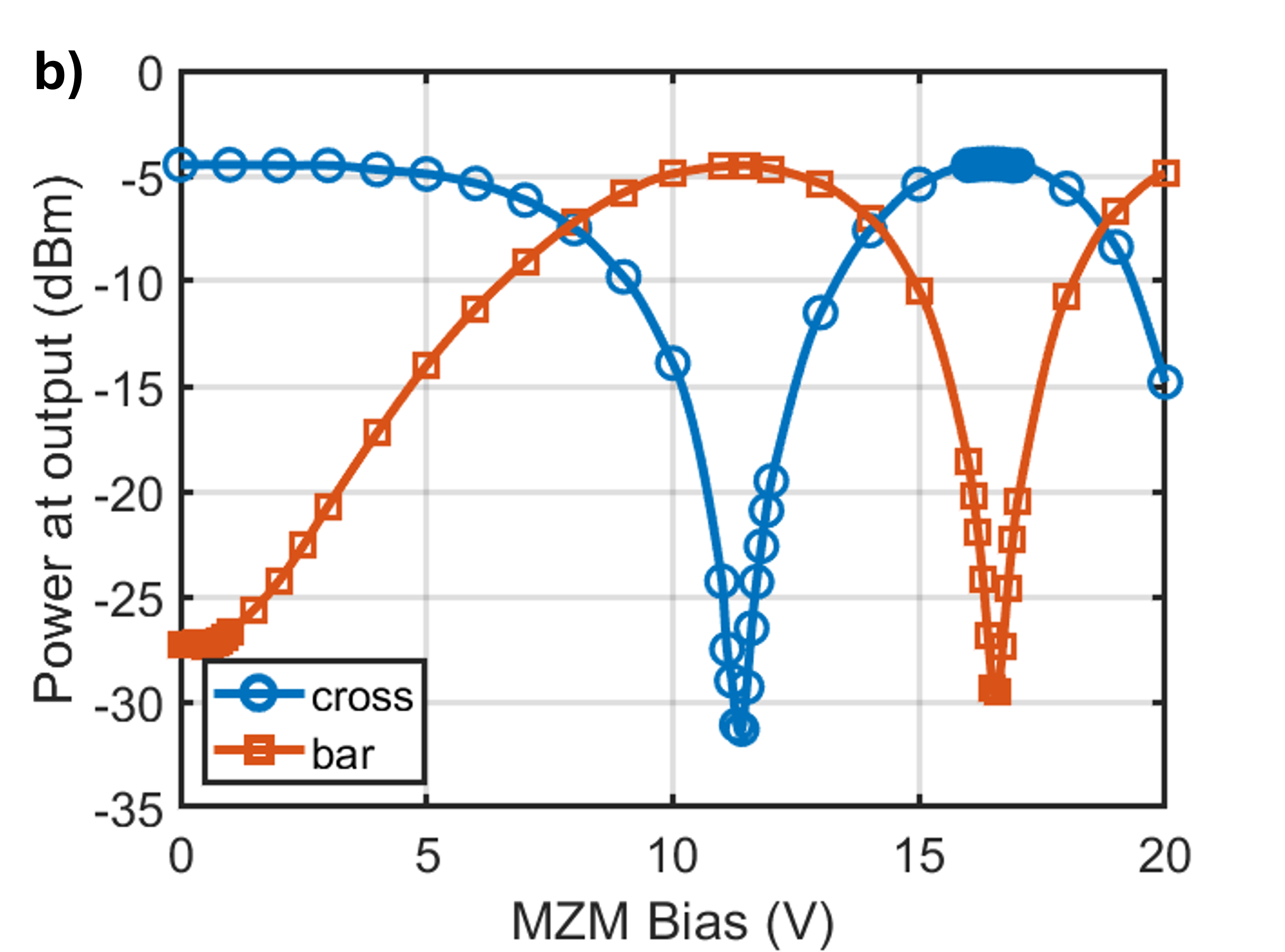}
        \phantomcaption
        \label{fig:MZM_Graphs_right}
     \end{subfigure}
    \caption{a) Total insertion loss as a function of different inputs and wavelengths. b) Measured interference of MZI 1.1 for bar and cross port when the external laser is coupled at inputs 2 and 1 respectively.}
    \label{fig:MZM_Graphs}
\end{figure}

A first set of measurements was carried out to assess the optical loss of the whole structure: the laser at 1550 nm was coupled to each input and the power at the output port was optimised by opportune control of the MZIs (See Supplementary Fig. 1). The wavelength was then swept from 1500 nm to 1600 nm in order to analyse the wavelength dependence of the device. The results are reported in the graph of figure~\ref{fig:MZM_Graphs_left}, where the insertion loss (IL) is reported as a function of the input port and wavelength. The IL is maximal for 1500~nm, and minimal at 1600~nm with average values across the ports of 19.7~dB and 13.8~dB, respectively. At the typical optical communication wavelength, 1550~nm, the mean IL value is 15.7~dB with a standard deviation of 1.1~dB. Despite the considerable size of the chip, the majority of the optical losses arise from the optical I/O and not from routing, as the former contributes around 5~dB/interface of loss and the latter introduces approximately $\sim$0.1~dB/cm at 1550~nm. The remaining 5.7~dB of losses are introduced by the interferometers, of which there are 5 for each input-output path, resulting in an average IL of $\sim$1.1~dB/MZI. The fact that losses decrease at higher wavelengths suggests that the MMIs are not at the correct working point at 1550~nm, likely as a result of a wider-than-designed central section.

%The laser at 1550 nm was coupled to each port and the power at the output port was optimised by opportune control of the MZIs. The wavelength was then swept from 1500 nm to 1600 nm in order to analyse the optical bandwidth of the device. The results are reported in figure \ref{fig:MZM_Graphs_left}. The insertion loss is maximal for 1500 nm, and minimal at 1600~nm with average values across the ports of -19.7~dB and -13.8~dB  respectively. At typical optical communication wavelength, 1550~nm, the mean value across the ports is -15.7~dB with a standard deviation of 1.1~dB.

A second set of measurements was carried out to evaluate the performance of the MZIs in terms of extinction ratio (ER), average power for a $\pi$ shift in the thermal tuners (P$_\pi$), thermal crosstalk, and phase shifter bandwidth. Figure~\ref{fig:MZM_Graphs_right} reports the interference fringes of MZI~1.1 as a function of applied bias voltage for its bar and cross port. 10~dBm of laser-light at 1550~nm was coupled individually to input 1 (cross) and input 2 (bar), and the consecutive MZIs were set to route all power to the output port. By sweeping the internal phase shifter voltage between 0 and 20~V, we evaluated an ER of 26.9~dB and 25.0~dB for the cross and bar states, respectively. The extinction ratio of all the MZIs in cross state was measured in a similar fashion, with an average value of 28.3~dB and a standard deviation of 2.4~dB. The performance of the MZIs is worse than expected by design: the IL should be lower, around 0.5~dB, and the ER should be around 40~dB. The $\pi$ shift power was then derived by evaluating the bias voltage corresponding to the minimum and maximum transmission and considering the resistance of each phase shifter. The average P$_\pi$ was measured to be 34.1~mW, with a standard deviation of 8.5~mW allowing for a total phase shift of more than $6\pi$ before failure. The P$_\pi$ values in PICs usually have a low variance as they depend on the heater resistance, which in our case should have been 1~k$\Omega$. Typically the P$_\pi$ of similar phase shifters is close to 25 mW~\cite{aalto2019open}. Here the excess and varying contact resistance of the heaters, with an average value of 4.1~k$\Omega$ and a standard deviation of 1.1~k$\Omega$, and thus power dissipation, was due to the use of the contact metallization process optimized for the PDs. This single metal process was used for the whole chip, while optimized multi-metallization steps should be considered in the future. The list of all measured resistances can be found in the supplementary material.

\begin{figure}[t]
    \centering
    \begin{subfigure}[b]{0.32\textwidth}
        \centering
        \includegraphics[width=\textwidth]{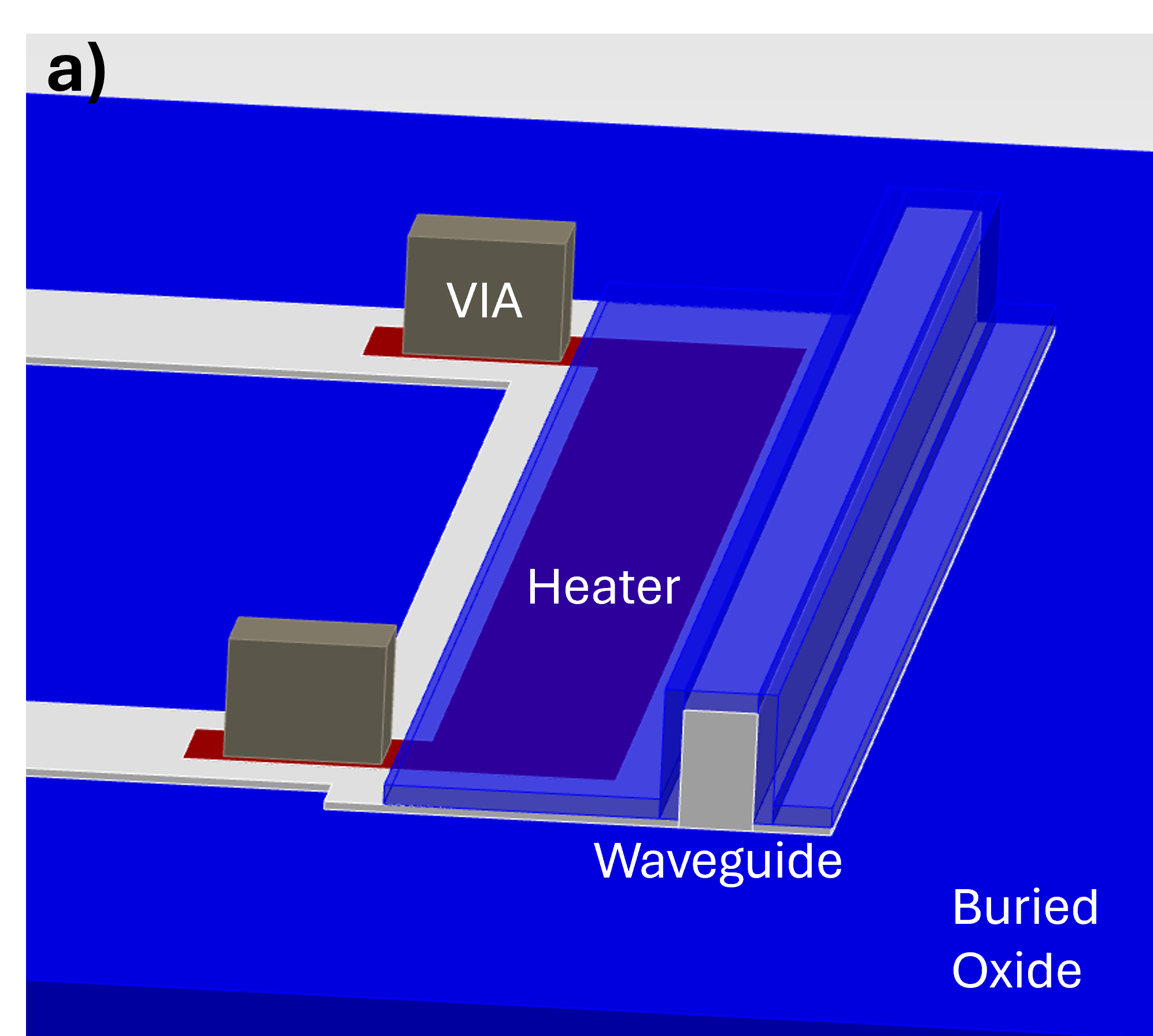}
        \phantomcaption
        \label{fig:SimSetup}
    \end{subfigure}
    \hfill
    \begin{subfigure}[b]{0.32\textwidth}
        \centering
        \includegraphics[width=\textwidth]{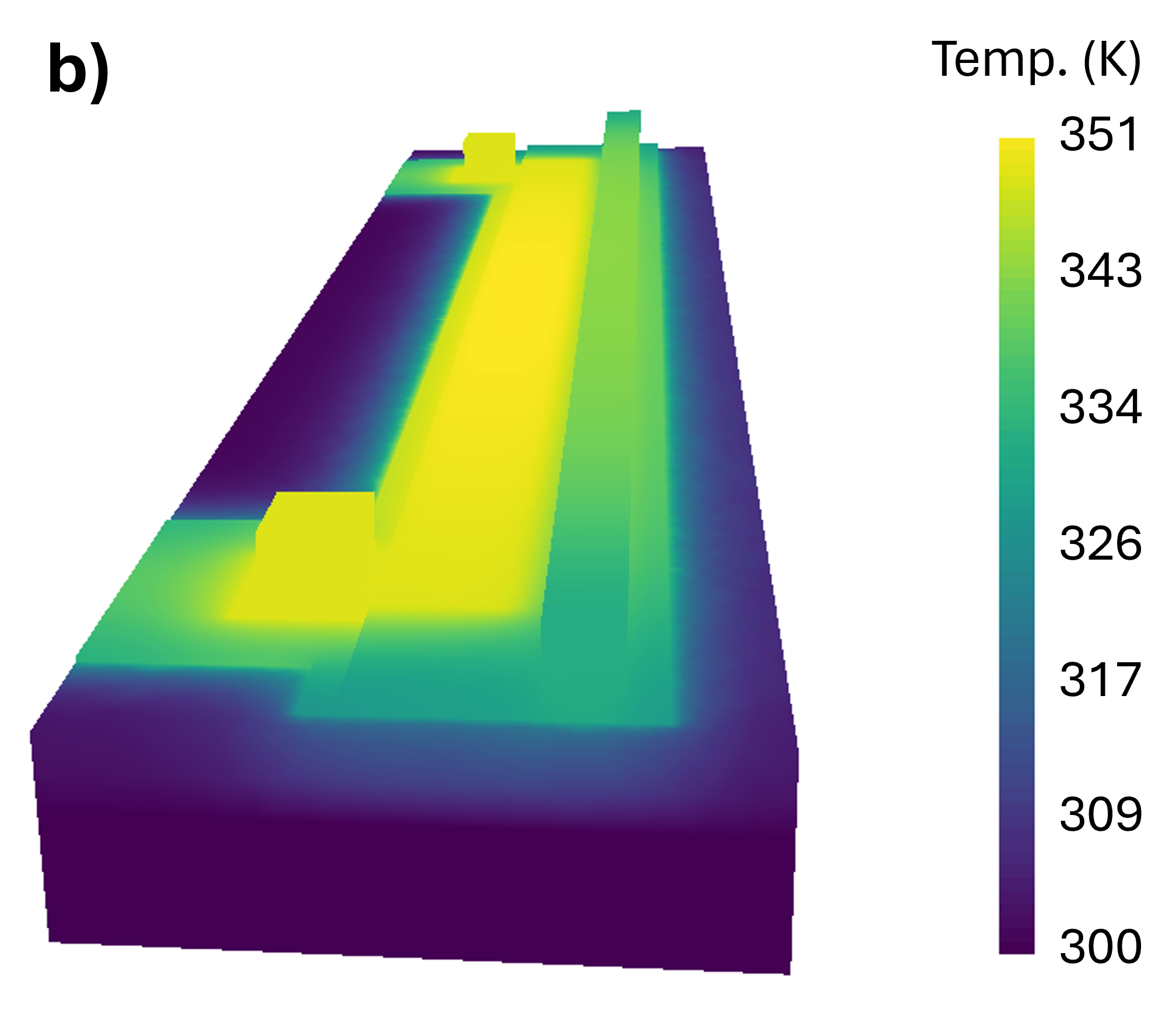}
        \phantomcaption
        \label{fig:ThermalGradient}
     \end{subfigure}
     \hfill
    \begin{subfigure}[b]{0.32\textwidth}
        \centering
        \includegraphics[width=\textwidth]{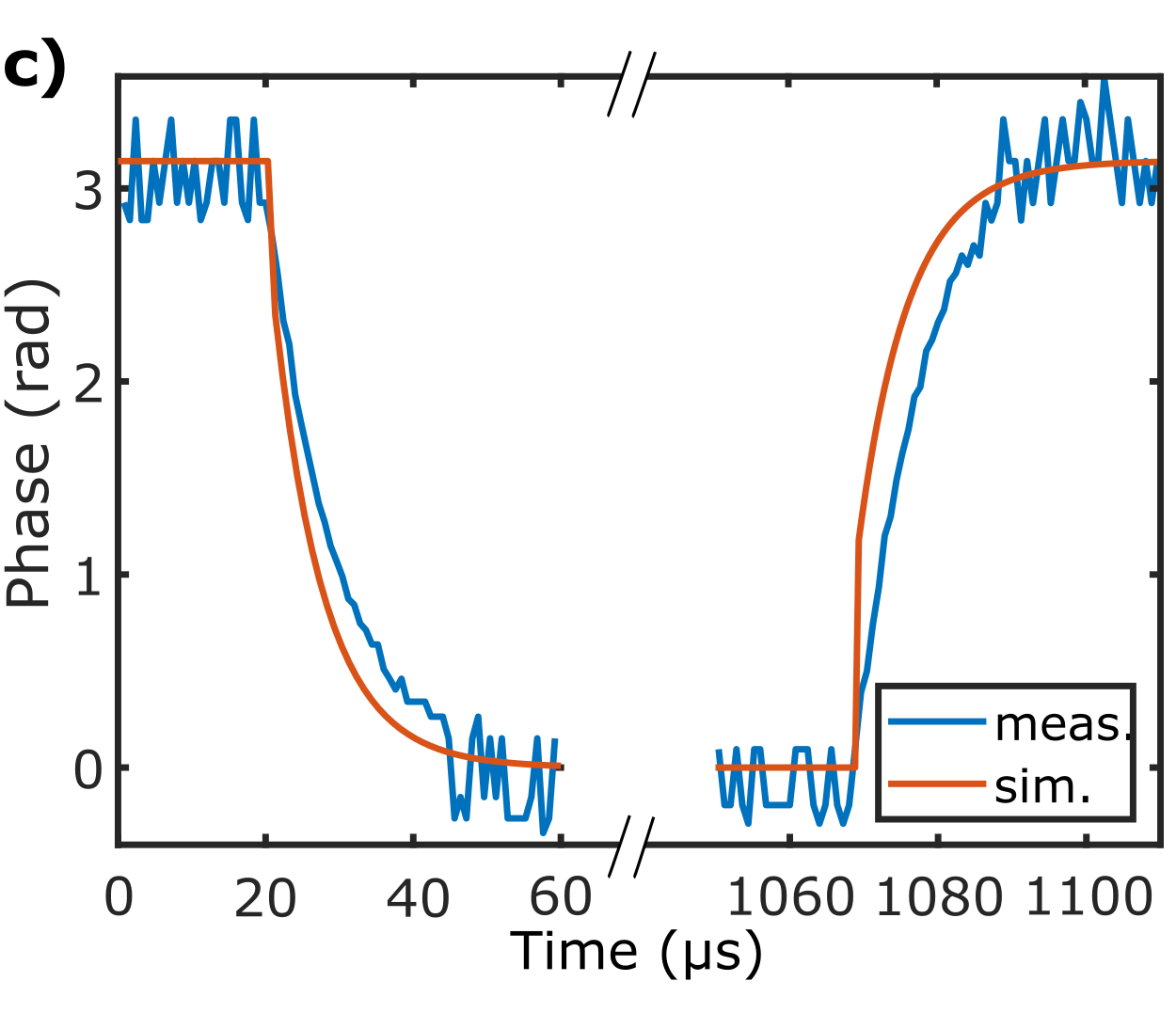}
        \phantomcaption
        \label{fig:Sim_risefall}
     \end{subfigure}
    \caption{a) Lumerical HEAT and CHARGE simulation setup for the implanted heater response; b) thermal distribution in steady state when a V$_\pi$ is applied to the shifter; c) comparison between the rise and fall signals for simulated and measured results.}
    \label{fig:Simulation}
\end{figure}

The bandwidth of the phase shifters is limited by the delay between the time the voltage signal is applied and when thermal equilibrium is reached. To characterise this limit the phase shifters were driven with a square wave signal, with a frequency of around 1~kHz and peak to peak amplitude equal to V$_\pi$, the voltage required to change the phase by $\pi$, producing an optical signal that goes from maximum to minimum at the output port. Before the measurements, we derived the bandwidth of our DAC, which was >300~kHz. The setup for this measurement is depicted in Supplementary Fig.~3. Before the fabrication of the circuit, the thermal phase shifters were numerically evaluated with Lumerical HEAT and CHARGE to predict the V$_\pi$ step response time. Figure~\ref{fig:SimSetup} represents the simulation setup, composed of the thick-silicon slab waveguide, passivation silica, the electrical vias, a 100~\si{\micro\metre} thermal shifter that runs parallel to the waveguide, and below, a buried oxide layer. A conductive and thermal transient simulation was performed to evaluate the rise and fall times when V$_\pi$ is applied to the electrical vias. Figure~\ref{fig:ThermalGradient} represents the thermal distribution at equilibrium when the heater is driven with V$_\pi$. The numerical analysis predicted a rise time of 7.81~\si{\micro\second} (128~kHz) and a fall time of 9.03~\si{\micro\second} (110~kHz). Figure~\ref{fig:Sim_risefall} compares the fall and rise signals, converted in relative phase shift, between the simulation and the experimental acquisition from the photodiode under actuation of the internal shifter of MZI 4.1.

\begin{figure}[t]
    \centering
    \begin{subfigure}[b]{0.49\textwidth}
        \centering
        \includegraphics[width=\textwidth]{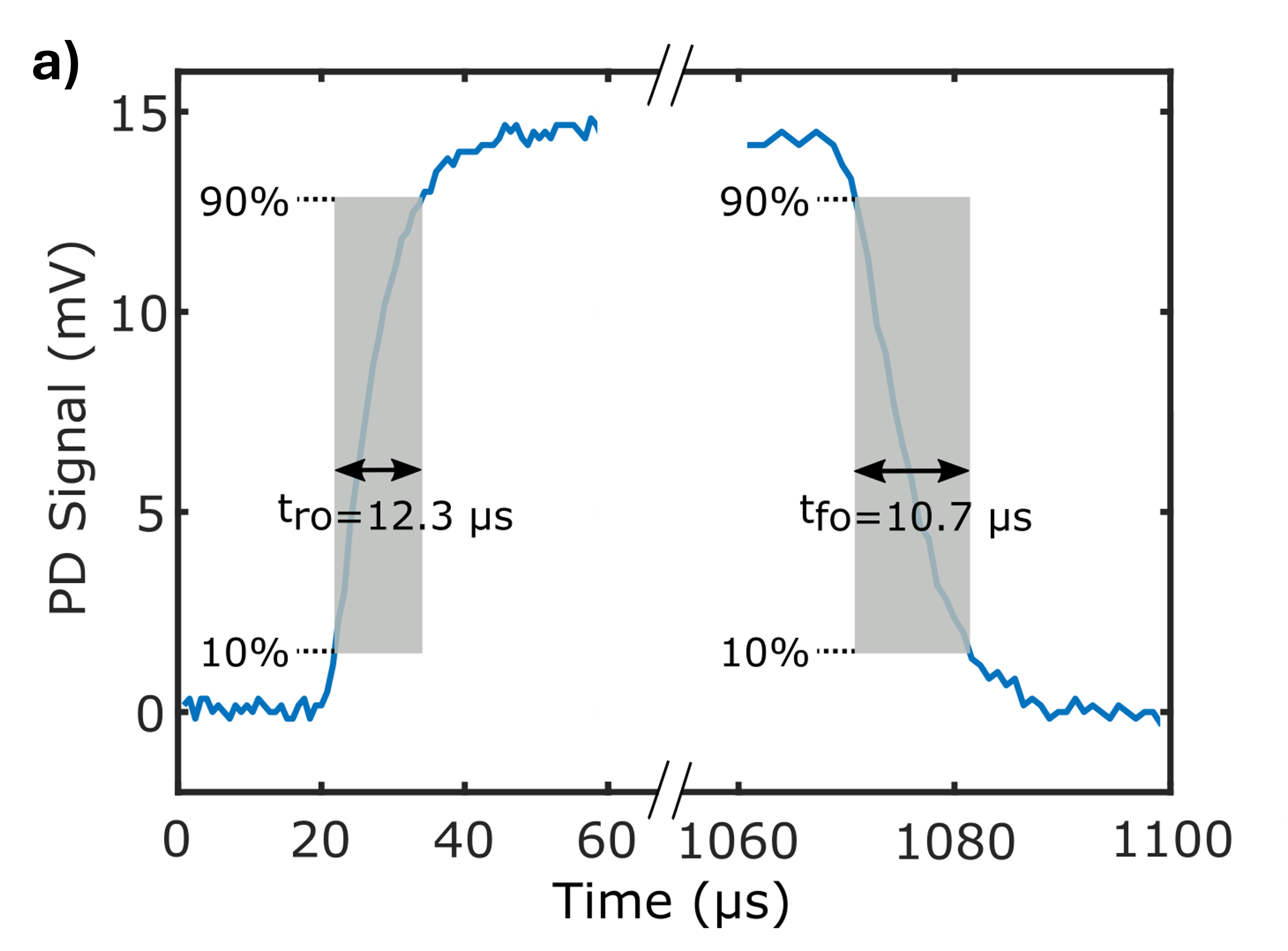}
        \phantomcaption
        \label{fig:MZM_risefall_left}
    \end{subfigure}
    \hfill
    \begin{subfigure}[b]{0.49\textwidth}
        \centering
        \includegraphics[width=\textwidth]{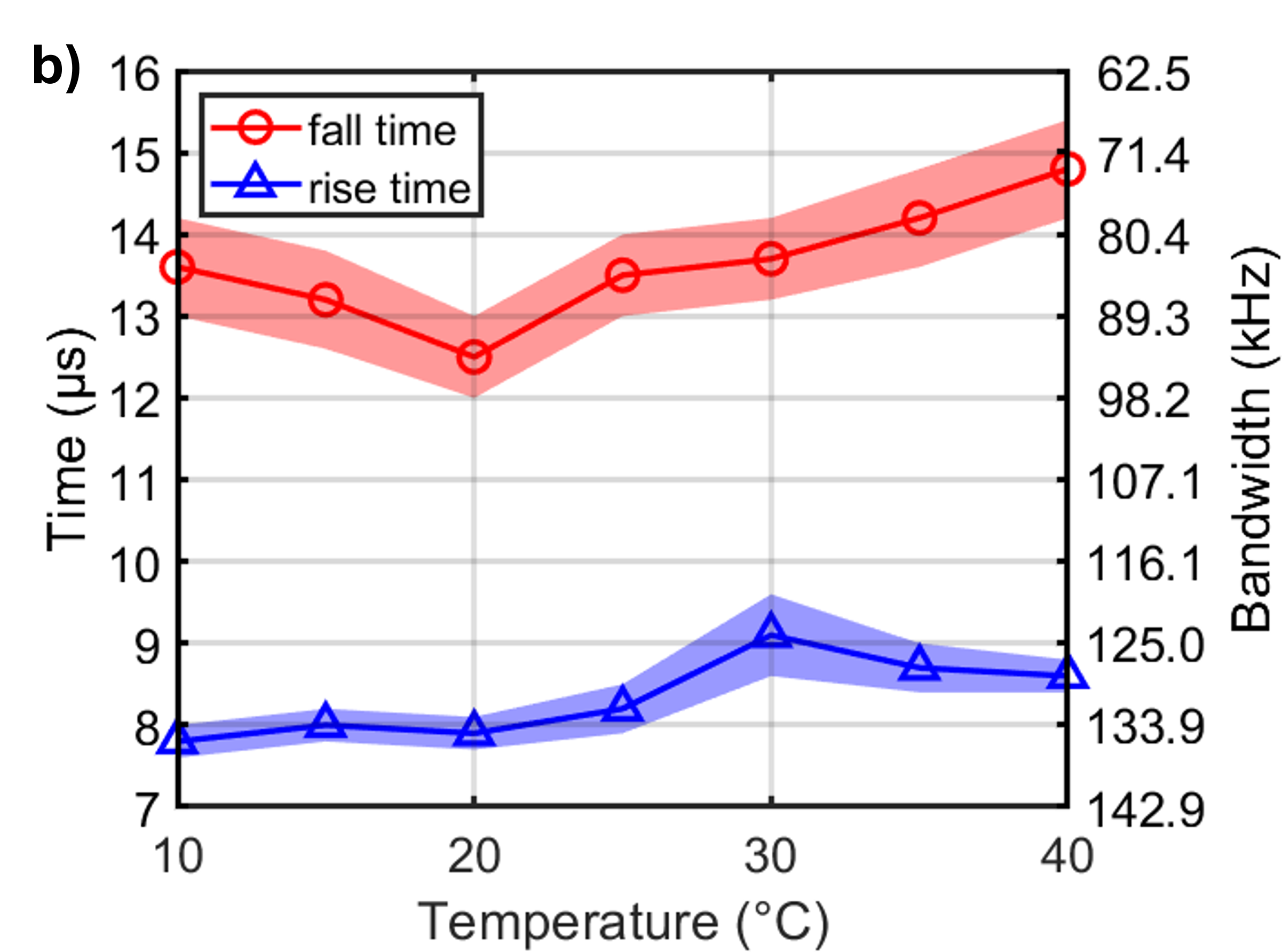}
        \phantomcaption
        \label{fig:MZM_risefall_right}
     \end{subfigure}
    \caption{a) Signal at PIC output during phase shifter modulation to show how rise time, $t_{fo}$, and fall time, $t_{ro}$, are calculated. b) Rise time, fall time and associated limits in bandwidth of MZI 1.1 as a function of temperature with shaded region representing $\pm$1 standard deviation.}
    \label{fig:MZM_risefallbandwidth}
\end{figure}

Figure~\ref{fig:MZM_risefall_left} illustrates how the rise and fall times are calculated for MZI~4.1. The symbols, $t_{ro}$ and $t_{fo}$, correspond to the optical rise and fall times respectively. Since the MZI is being driven in cross state, $t_{fo}$ corresponds to when the phase shifter is heating (rise time) and $t_{ro}$ when the phase shifter is not activated and the device is returning to room temperature (fall time). In this instance, $t_{ro}$=1.23~$\times$~10$^{-5}$ s and $t_{fo}$=1.07$\times$~10$^{-5}$ s corresponding to respective limits in bandwidth of 81 kHz and 94 kHz. Four phase shifters were characterised in this way, the mean values of rise and fall time were calculated over several modulation cycles, resulting to be 12.4~\SI{}{\micro\second} (80.6~kHz) and 9.4~\SI{}{\micro\second} (106~kHz), respectively.

Since the thermalisation time is dependent on the temperature of the PIC, for MZI 1.1, a characterization was performed in a range of temperatures from 10~$^{\circ}$C to 40~$^{\circ}$C, to simulate the operation of the PIC in different environmental conditions. Figure~\ref{fig:MZM_risefall_right} reports the results of this investigation, showing a notable increase in the fall time, as expected from the reduced temperature gradient, when the heater is switched off. These phase shifters are fast enough to use in a coherent combination context, for which a 50 kHz modulation at small amplitude ($\leq$~1~radians) was proposed~\cite{lucas2024spatial}. The thermal cross talk effect between the internal shifter of MZI 1.1 and the surrounding phase shifters was investigated by coupling light into port 1, setting the voltage at the internal shifter such that the power at the output port was at a minima (more sensitive to changes in temperature), and measuring the output power for both 0~V and 20~V applied to the closest surrounding phase shifters (corresponding to >2$\pi$ phase change). The maximum observed change in output power resulting from the cross talk was 2.1~dB corresponding to a change in phase of 0.074 radians, when the external phase shifter of MZI 1.1 was heated, it was less than 0.3~dB (0.031 radians) for all other cases. The very low thermal crosstalk is ensured by the deep trenches around the heater and waveguides of the VTT Silicon process~\cite{aalto2019open}.

\begin{table}[h]
    \begin{subtable}[h]{0.49\textwidth}
    \centering

    \begin{tabular}{|c|c|c|c|}
    \hline
    \textbf{Temp}&\multicolumn{3}{|c|}{\textbf{Dark Current (nA)}} \\
    \cline{2-4}
    \textbf{($^{\circ}$C)} &\textbf{0V} & \textbf{-1V}& \textbf{-2V} \\
    \hline

    \hline
    10&1.9 & 496.4  & 714.1   \\
    \hline
    15&1.6 & 633.4  & 906.0   \\
    \hline
    20&1.5 & 791.8  & 1125.5  \\
    \hline
    25&1.6 & 1004.6 & 1426.0  \\
    \hline
    30&1.5 & 1258.3 & 1788.2   \\
    \hline
    35&1.7 & 1553.0 & 2236.1  \\
    \hline
    40&1.7 & 1932.2 & 2794.7  \\
    \hline
    \end{tabular}
    \caption{}
    \label{tab:darkcurrenttemp}
    \end{subtable}
    \hfill
    \begin{subtable}[h]{0.49\textwidth}
    
    \centering
    
    \begin{tabular}{|c|c|c|c|c|c|}
    \hline
    \textbf{Photodiode}&\multicolumn{5}{|c|}{\textbf{Bandwidth (GHz)}} \\
    \cline{2-6}
    \textbf{} &\textbf{0.0V} & \textbf{-0.5V}& \textbf{-1.0V}& \textbf{-1.5V} &\textbf{-2.0V}\\
    \hline

    \hline

    1.1& 2.6 & 9.1  & 11.8 & 13.3 & 14.1  \\ \hline
    2.1&2.6 & 7.5  & 10.2 & 11.8 & 13.2  \\ \hline
    4.2&4.4 & 8.6  & 11.8 & 13.3 & 14.2 \\ \hline
    1.16&2.1 & 7.5  & 11.8 & 13.6 & 15.0 \\ \hline

    \end{tabular}
    \caption{}
    \label{tab:bandwidth}
    
    \end{subtable}
\caption{a) Dark currents for photodiode 2.3 as a function of bias and temperature and bias. b) 3~dB bandwidth for different photodiodes as a function of bias.}
\end{table}

\begin{figure}[h!]
    \centering
    \begin{subfigure}[b]{0.49\textwidth}
        \centering
        \includegraphics[width=\textwidth]{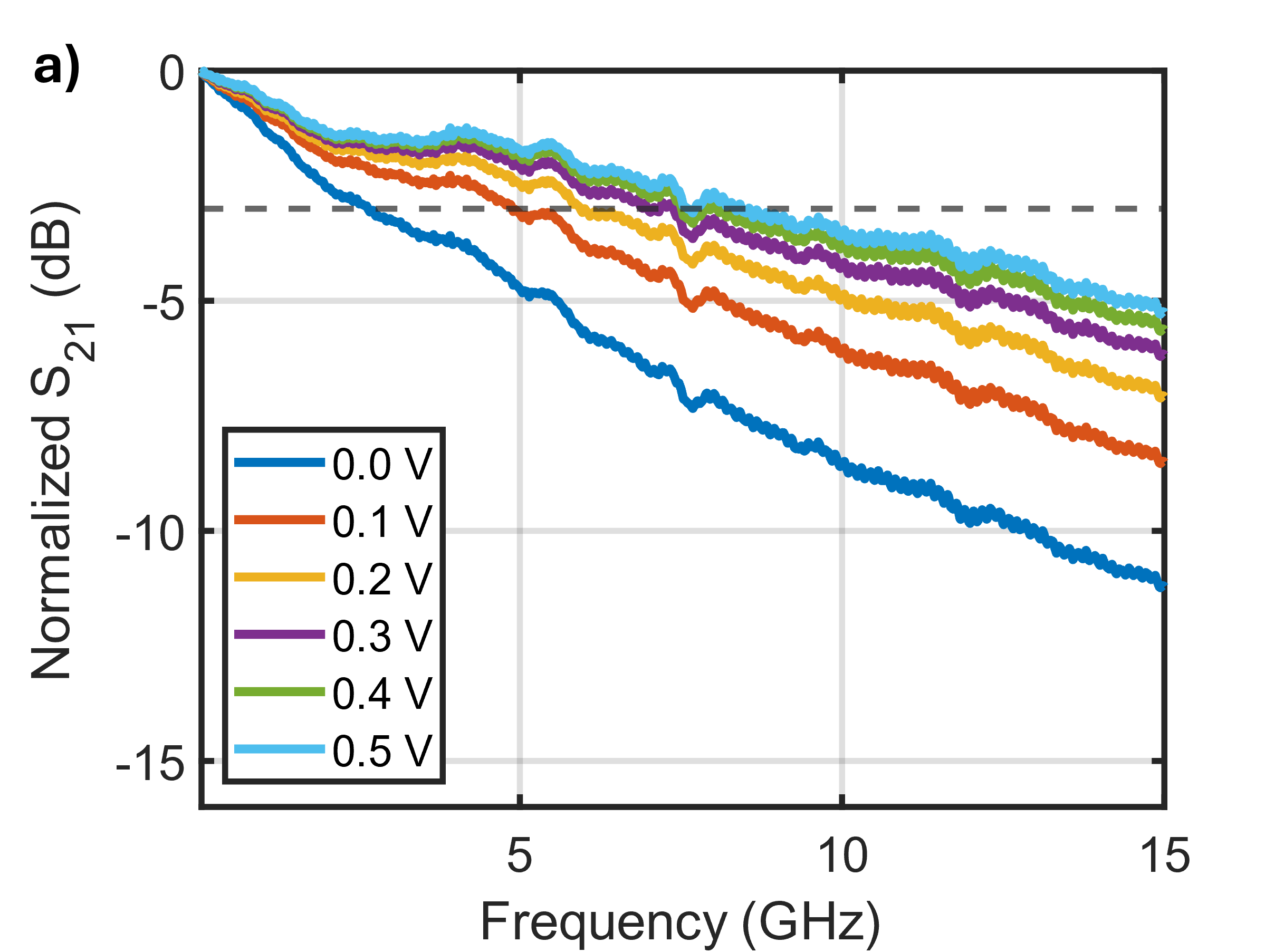}
    \end{subfigure}
    \hfill
    \begin{subfigure}[b]{0.49\textwidth}
        \centering
        \includegraphics[width=\textwidth]{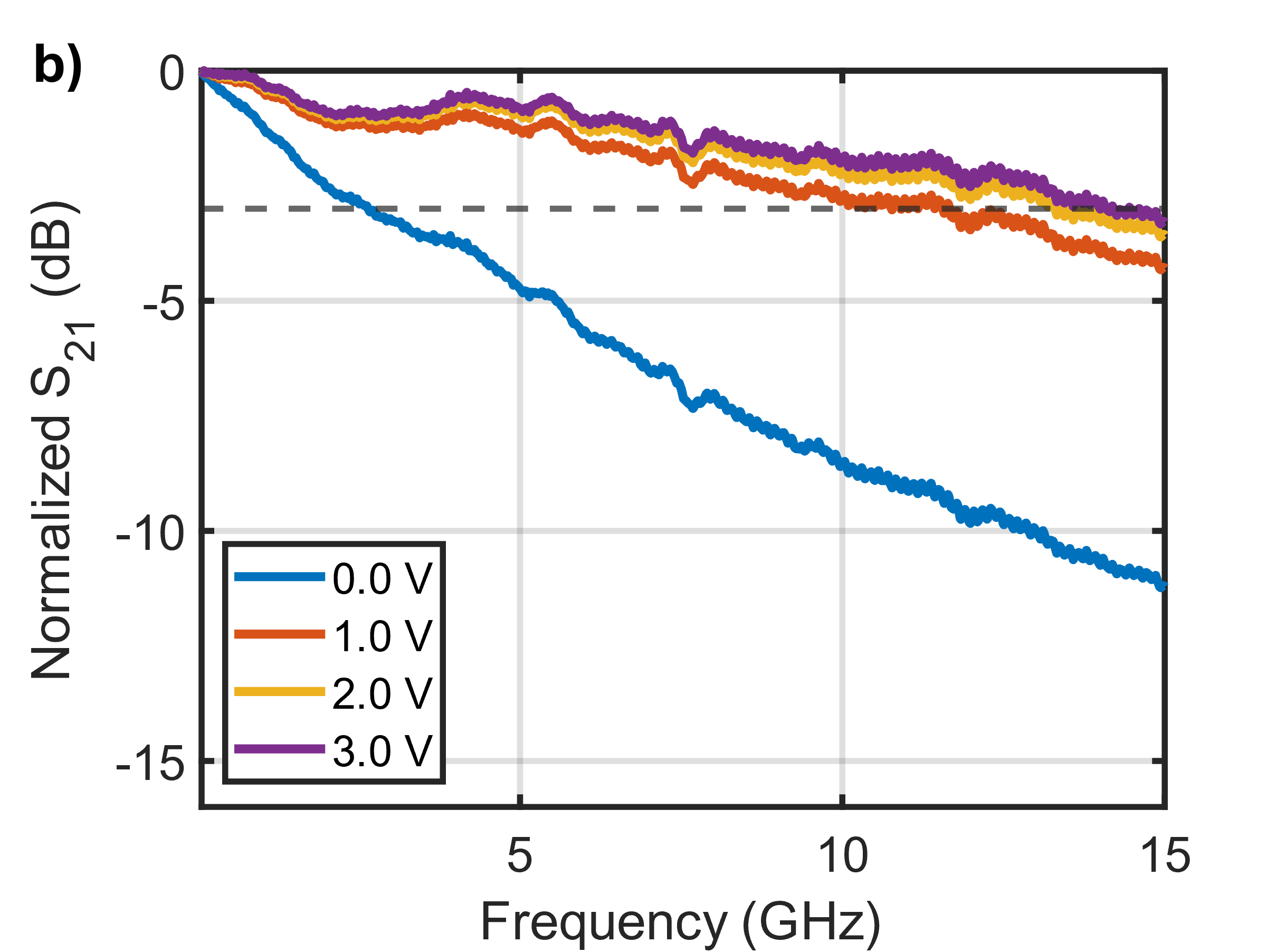}
     \end{subfigure}
    \caption{Normalized S$_{21}$ response of photodiode 2.1 for a range in reverse bias a) 0.0 V - 0.5 V b) 0.0 V - 3.0 V. The horizontal grey dashed line corresponds to -3~dB.}
    \label{fig:PD_BW}
\end{figure}

A final set of measurements was carried out to assess the germanium photodiode performance in terms of dark current, responsivity, and optical bandwidth. Without any optical input, the dark current was measured as a function of bias for three different photodiodes (1.7, 1.16, 2.3), the average values at 0 V, 1 V and 2 V of negative bias are 1.1~$\pm$~0.3~nA, 764~$\pm$~74~nA and 1096~$\pm$~128~nA respectively. For photodiode 2.3, the dark current was also measured as a function of both bias and temperature with results reported in table~\ref{tab:darkcurrenttemp}, when negative bias was applied there was around four times as much dark current at a temperature of 40~$^{\circ}$C compared to 10~$^{\circ}$C. The responsivity of the photodiodes was estimated $\approx$~0.8~A/W in DC by monitoring the output current when 10~dBm power of laser was injected, considering the coupling and propagation losses. The measurement of the 3~dB bandwidth of the monitor PDs could not be performed using the packaged PIC, as these were wirebonded to DC connectors. The PDs were characterised on a naked chip fabricated in the same run and wafer at a probe station. The laser was coupled to a external modulator with >25 GHz bandwidth, whose output was divided by a 3-dB splitter with one branch leading to a 50~GHz PD with a known bandwidth response, and the other to a SMF fiber which was edge coupled to the optical inputs on the chip. A microwave vector network analyzer was used with the output port coupled to the modulator input port, one receiver port was connected to the known PD and the other to a RF probe used to contact the PDs on the chip. This setup is represented in Supplementary Fig.~4. Different negative biases were supplied with a bias-tee at the RF probe, with the DC port connected to a source measure unit. With this configuration we subtracted the response of the known PD branch from the S$_{21}$ response of the device under test, to compensate for the response of the RF equipment common to both outputs. The S$_{21}$ frequency response of the known PD itself and the RF tip were then removed, and with a separate characterization, a correction was applied for the difference in output signal paths (only one had a bias-tee).  With this method, the 3~dB bandwidths at differing negative bias were characterised for 4 photodiodes selected in different areas of the PIC; Table~\ref{tab:bandwidth} reports these results. An example of a typical PD response, in this case for PD 2.1, as a function of applied negative bias is shown in Fig.~\ref{fig:PD_BW}, where the dashed horizontal line corresponds to -3~dB. The germanium photodetectors showed a large bandwidth increase by applying the first 0.5~V of negative bias, (from 2.6 to 7.5~GHz for PD 2.1), with a less pronounced improvement as the negative bias was increased. As the monitor photodiodes for the adaptive optics algorithm read signals at some tens of kHz, zero bias may be used as it provides sufficient bandwidth and a very low dark current (on the order of few nA), that would otherwise increase significantly.

\section*{Discussion}
\label{sec:discussion}

The device thermal phase shifters surpassed the target bandwidth of 50~kHz, while providing a sufficient number of $\pi$ shifts (>~6$\pi$) to be used with an appropriate control algorithm in real time. The main shortcoming which prevents the use of the presented coherent combiner in realistic applications resides in the losses, which resulted much higher than expected, as discussed in the previous section. The main contribution resides in the optical I/O, accounting for 5~dB per facet, with the MZIs also contributing, whose constituent MMI were fabricated wider than designed, leading to an overall average loss of 15.7~dB at 1550~nm. However, the losses can be considerably reduced in future fabrications through the addition of the optical interposer chip inserted between the SMFs and the PIC, which lowers the coupling losses to 1.5~dB per facet. Moreover, the current VTT process has waveguides with improved smoothness thanks to a hydrogen-annealing step, thus reducing the propagation losses with a demonstrated value of just 0.027~dB/cm, a more than threefold improvement with respect to the previous value of 0.1 dB/cm~\cite{Marin:23}. Finally, the IL of each MZI can be lowered to $\leq$0.5~dB/MZI with a correct fabrication of the constituent MMIs~\cite{aalto2019open}. With this performance, the same chip with 32 inputs would have a total loss of 5.5~dB, accounting for 2 optical I/O and 5 MZI encountered by each path, and disregarding propagation losses. In the case of a 64-input device, the losses increase by 0.5~dB as there is one additional MZI in all input-output paths. 

Aside from improvements in the losses, the PIC power consumption can be decreased considerably: the thermal phase shifter contact resistance should be lower, granting a P$_\pi$ $\leq$~25~mW, instead of the 34.1~mW found with the current chip. Moreover, the heater thermal isolation can be further improved by removing the silicon substrate underneath the heater, leading to a P$_\pi$ on the order of 1~mW~\cite{aalto2019open}. Table~\ref{tab:compare} reports a comparison between our work and other coherent combiner PICs from the literature. In the last row, we have reported the prospective performance of a 32/64-input coherent combiner according to the latest demonstrated performance of the VTT process and introduction of an interposer, as discussed above. We omitted the work of Melloni et al.~\cite{melloni2023programmable} for the lack of information reported about the PIC (which has 16 inputs with integrated photodetectors). The table reports number of supported modes (inputs), footprint, whether the PIC has integrated PDs, average total loss, phase shifter rise/fall time, P$_\pi$, and MZI IL. Although our chip has the highest optical losses at telecom wavelength, it demonstrated good performance in terms of MZI IL, phase shifter rise/fall time, and the best in term of power consumption. The prospective device would have the lowest losses even with a 64 input configuration, while also improving the power consumption. As discussed by L. Rinaldi~\cite{lucaThesis}, a 32-input chip would be suitable for medium-small turbulence, as in horizontal links and LEO satellites close to the zenith, while a device with 64 inputs would also allow the mitigation of strong turbulence conditions, as for LEO satellites close to the horizon.

\begin{table}[h]
\centering
    \begin{tabular}{c|ccccc}
    \toprule
        \textbf{PIC Combiner} & \textbf{Billaud \emph{et al.}~\cite{Billaud2022Turbulence}}          & \textbf{Billault \emph{et al.}~\cite{Billault:21}}  & \textbf{Billaud \emph{et al.}~\cite{Billaud45mode2024}} & \textbf{This work}                   & \textbf{Prospective} \\ \hline
        \textbf{\# inputs}              & 8     & 15    & 45    & 32        & 32 / 64 \\ 
        \textbf{Platform}               & SOI   & SOI   & SOI   & Thick SOI & Thick SOI \\ 
        \textbf{Footprint (cm$^2$)}              & $\approx$ 1 & NA & NA        & 0.5 & 0.5 / 1.0 \\ 
        \textbf{Integrated PDs}         & Yes   & Yes  & No    &   Yes     & Yes \\
        \textbf{Avg losses (dB) }    & 9.8 @ 1590 nm     & 7-11 @ 1535 nm   & 9 dB & 15.7 @ 1550 nm        & 5.5 / 6.0 \\ 
        \textbf{Rise time} (\SI{}{{\micro\second}})     & 10    & $\approx$ 25  & 5.1 & 12.4 & $\approx$10 \\ 
        \textbf{Fall time} (\SI{}{{\micro\second}})     & 7.6                           & $\approx$ 25      & NA      & 9.4                         & $\approx$10         \\ 
        \textbf{P$_\pi$ (mW)}                & \textless{}40              & NA         & NA        & 34.1                     & $\leq$25         \\ 
        \textbf{MZI IL (dB)}           & $\approx$ 1                      & 0.6        & NA        & 1.1                     & $\leq$0.5         \\ \bottomrule
    \end{tabular}  
\caption{Comparison of coherent combiners on PICs from literature with the presented work and a prospective device.}
\label{tab:compare}
\end{table}

To further reduce the losses, the combiner architecture can be changed to enable phase-only correction of the decomposed modes~\cite{lucas2024spatial}. This can be easily done by substituting the MZIs in the binary tree architecture with 3-dB (50/50) 2$\times$2 directional couplers, or equivalently MMIs. Here, the phase control is implemented by two thermal tuners at the inputs of each element, while the feedback for the control loop is collected by a PD at each coupler output. For the VTT process, the MMI IL is $\leq$~0.2~dB instead of the $\approx$~0.5~dB for an MZI~\cite{aalto2019open}, so accounting for 1.5~dB per I/O interface, a 32-input coherent combiner would have an overall loss of 4~dB, and a 64-input combiner would have 4.2~dB of loss, where propagation losses are neglected as extremely low (0.027~dB/cm with hydrogen-annealing). However, a phase-only correction of the input modes is equivalent to the MZI scheme only if the modes have the same amplitude. In a realistic scenario, the amplitudes can be modelled as independent random variables following a Rayleigh distribution. For this reason, as discussed by L. Rinaldi~\cite{lucaThesis}, for a phase-only device a higher number of modes is required to obtain an equivalent correction: around 25 for medium-small turbulence, and more than a hundred for strong turbulence. These numbers can be reduced to 16 and 64 respectively, if a tip-tilt correction is implemented in the receiving chain before mode decomposition~\cite{lucaThesis}.

\begin{comment}

\begin{table*}[h!]

\begin{center}
\begin{tabular}{|c|c|c|c|c|c|}
\hline
\textbf{Photodiode}&\multicolumn{5}{|c|}{\textbf{Bandwidth (GHz)}} \\
\cline{2-6}
\textbf{} &\textbf{0.0V} & \textbf{-0.5V}& \textbf{-1.0V}& \textbf{-1.5V} &\textbf{-2.0V}\\
\hline

\hline

1.1& 2.6 & 9.1  & 11.8 & 13.3 & 14.1  \\ \hline
2.1&2.6 & 7.5  & 10.2 & 11.8 & 13.2  \\ \hline
4.2&4.4 & 8.6  & 11.8 & 13.3 & 14.2 \\ \hline
1.16&2.1 & 7.5  & 11.8 & 13.6 & 15.0 \\ \hline

\end{tabular}
\caption{3 dB bandwidth for different photodiodes as a function of bias.}
\label{tab4}
\end{center}
\end{table*}
\end{comment}

\begin{comment}
\begin{table}[h!]
\caption{3 dB bandwidth for photodiode 2.1 as a function of bias.}
\begin{center}
\begin{tabular}{|c|c|c|c|c|c|c|}
\hline
\textbf{}&\multicolumn{6}{|c|}{\textbf{Bias (V)}} \\
\cline{2-7}
\textbf{} &\textbf{0.0} & \textbf{-0.1}& \textbf{-0.2}& \textbf{-0.3} &\textbf{-0.4}&\textbf{-0.5}\\
\hline

\hline

Bandwidth (GHz)&2.6&4.9 & 5.9  & 6.9 & 7.5 & 7.6 \\ \hline

\hline

\end{tabular}
\label{tab4}
\end{center}
\end{table}

\end{comment}

\section*{Conclusions}
\label{sec:conclusions}

Free space optical links enable wireless communication at high data rates, and are envisaged as a key component of future communication systems, spanning 5G/6G, quantum links, and satellite communications. However, atmospheric turbulence introduces distortions to the beam which limits the effective use of FSO in horizontal and satellite links.

Here, we have reported the design, fabrication, and characterization of an integrated coherent optical combiner for turbulence mitigation in free space optical links. The PIC was fabricated with the Thick SOI open platform offered by VTT, featuring 32 inputs and containing 31 integrated MZIs and photodetectors. The phase shifters were found to have a bandwidth of 81~kHz, an average P$_\pi$ of 34.1~mW, and are able to shift the phase by more than 6$\pi$, meeting the requirements for turbulence correction in horizontal and LEO-to-ground links. Due to fabrication problems, the chip losses were much higher than expected, with an average value of 15.7~dB at a wavelength of 1550~nm. This is the main shortcoming preventing its use in real scenarios.

Nonetheless, by leveraging an optical interposer and the reduced losses offered by the new VTT process, a prospective device with the same number of inputs would reduce the losses to just 5.5~dB. For a 64 input device, the losses would increase to 6 dB, lower than any fabricated PIC combiner in literature. These numbers can be further reduced to 4~dB and 4.2~dB respectively if the device were to implement a phase-only correction.

%Moving forwards, a device with acceptable losses and a larger number of inputs will be designed, required for dealing with strong turbulence conditions (satellite close to horizon) in LEO-ground links. Indeed, by leveraging an optical interposer and the reduced losses offered by the updated VTT process, a prospective device based on the same architecture with 64 inputs would have an insertion loss of 6 dB, lower than any fabricated PIC combiner in literature. The power consumption of the phase shifters could also be improved for a P$_\pi$ $\leq$~25~mW. The losses could be further reduced to 4.2~dB if the device were to implement a phase-only correction (requiring an alternative design), where the loss of a degree of freedom of the modulators would result in a higher number of required input modes to perform an equivalent correction. The introduction of a tip-tilt correction in the receiver apparatus would nonetheless limit the number of required modes to 64, even in strong turbulence conditions. 

\bibliography{sample,ref_vm}

\section*{Acknowledgements}

The authors would like to thank Dr. L. Rinaldi for fruitful discussions about his work on atmospheric turbulence mitigation. This work was partially supported by the EU Horizon 2020 Programme through the Actphast 4R Project under Grant Agreement n° 779472. The authors acknowledge several of VTT's experts in Si photonics fabrication and PIC packaging for implementing the prototype device. VTT's Si photonics activities are part of the Research Council of Finland Flagship Programme, Photonics Research and Innovation (PREIN), decision number 346545. 

\section*{Author contributions statement}

Conceptualization V.M. and G.C.; methodology L.D.M., V.M., G.C., and L.K.; circuit design and simulation L.D.M.; chip fabrication and packaging M.C. and M.K.; photonic chip characterization L.D.M., Y.L. and P.S.K.; data analysis, L.D.M., Y.L. and P.S.K.; writing—original draft preparation, L.D.M., V.M. and P.S.K.; supervision, G.C.; funding acquisition, G.C., V.M., M.C.; All authors contributed to the writing and review of the manuscript.

\section*{Competing interests}
The authors declare no competing interests.

\end{document}